\documentclass[journal=jacsat,manuscript=article]{achemso}

\usepackage[version=3]{mhchem} 
\usepackage{graphicx}
\usepackage{dcolumn}
\usepackage{bm}

\usepackage{cleveref}
\usepackage[utf8]{inputenc}
\usepackage[T1]{fontenc}
\usepackage{mathptmx}
\usepackage{etoolbox}
\usepackage{siunitx}
\usepackage{graphicx}
\usepackage{dcolumn}
\usepackage{bm}
\usepackage{xcolor}
\usepackage{verbatim}
\usepackage{soul}
\usepackage{amsmath,amssymb}

\author{A. Greco}
\email{angelo.greco@nano.cnr.it}
\affiliation{NEST, Istituto Nanoscienze-CNR and Scuola Normale Superiore, I-56127 Pisa, Italy}

\author{Q. Pichard}
\affiliation{NEST, Istituto Nanoscienze-CNR and Scuola Normale Superiore, I-56127 Pisa, Italy}

\author{F. Giazotto}
\email{francesco.giazotto@sns.it}
\affiliation{NEST, Istituto Nanoscienze-CNR and Scuola Normale Superiore, I-56127 Pisa, Italy}

\title[An \textsf{achemso} demo]
  {Josephson diode effect in monolithic dc-SQUIDs based on 3D Dayem nanobridges}

\abbreviations{IR,NMR,UV}

\begin{document}

\begin{tocentry}

Some journals require a graphical entry for the Table of Contents.
This should be laid out ``print ready'' so that the sizing of the
text is correct.

Inside the \texttt{tocentry} environment, the font used is Helvetica
8\,pt, as required by \emph{Journal of the American Chemical
Society}.

The surrounding frame is 9\,cm by 3.5\,cm, which is the maximum
permitted for  \emph{Journal of the American Chemical Society}
graphical table of content entries. The box will not resize if the
content is too big: instead it will overflow the edge of the box.

This box and the associated title will always be printed on a
separate page at the end of the document.

\end{tocentry}

\begin{abstract}
It was recently  experimentally proved that the superconducting counterpart of a diode, i.e., a device that realizes nonreciprocal Cooper pairs transport, can be realized by breaking the spatial and time-reversal symmetry of a system simultaneously. Here we report the theory, fabrication, and operation of a monolithic dc superconducting quantum interference device (dc-SQUID) that embedding three-dimensional (3D) Dayem nanobridges as weak links realizes an efficient and magnetic flux-tunable supercurrent diode. The device is entirely realized in Al and achieves a maximum rectification efficiency of $\sim 20\%$, which stems from  the high harmonic content of its current-to-phase relation only without the need of any sizable screening current caused by a finite loop inductance. Our interferometer can be easily integrated with state-of-the-art superconducting electronics, and since it does not require a finite loop inductance to provide large rectification its downsizing is not limited by the geometrical constraints of the superconducting ring.
\end{abstract}


Superconducting electronics is superior in terms of speed and power consumption with respect to classical silicon-based electronics due to its vanishing resistance, that eliminates ohmic losses and increases RC time constants. This fact, together with the possibility of integrating superconducting elements with quantum computers make this an appealing technology not only for low energy impact purposes, but also as a possible tool to boost up superconducting quantum technologies. Nevertheless, superconducting electronics suffer from the lack of dissipationless analog components for the processing and shaping of signals. In classical electronics, we can rely on semiconducting PN junctions as nonreciprocal elements, which are elementary building blocks for several fundamental devices like transistors and diodes. It has been demonstrated the possibility of achieving nonreciprocal  transport in superconducting systems \cite{akatsuni2017,wakatsuki2017}, practically realizing what is called a \textit{supercurrent diode} \cite{jiang2022,bauriedl2022,ando2020}. The latter is typically intended as a two-terminal device that can carry dissipationless currents of different absolute values between the two leads, depending on the orientation of the charge carriers motion. This effect has been observed in many different types of superconducting systems embedding Josephson junctions realized with hybrid structures \cite{gupta2023, wu2022,turini2022josephson} or ferromagnetic layers \cite{aladyshkin2009}. Due to the presence of hybrid and ferromagnetic components, which are nowadays non-standard materials in state of art superconducting electronics, these realizations of the supercurrent diode effect (SDE) are challenging to be reliably embedded in large-scale systems. The SDE has been observed also in systems based on low-$T_c$ conventional superconductors, where the spatial inversion symmetry is broken by means of geometrical features \cite{lyu2021,satchell2023}. When the SDE is realized by means of the Josephson coupling we refer to it as the Josephson diode effect (JDE) \cite{zhang2022,baumgartner2022}. The JDE can be obtained also via asymmetric superconducting quantum interference devices (dc-SQUIDs) where the screening effect of the supercurrents is not negligible \cite{fulton1972}, or where the harmonic content of the weak links is tunable on-demand \cite{souto2022, paolucci2023}. These realizations, which can be obtained with conventional Al or Nb technology \cite{levensonfalk2011, bouchiat2001,margineda2023sign} can be more easily integrated with many fabrication processes.\\
Here, we show a sizable rectification efficiency achievable in a dc-SQUID where the Josephson effect is obtained through two three-dimensional (3D) nanoconstrictions (i.e., 3D Dayem nanobridges).
\begin{figure}
\includegraphics{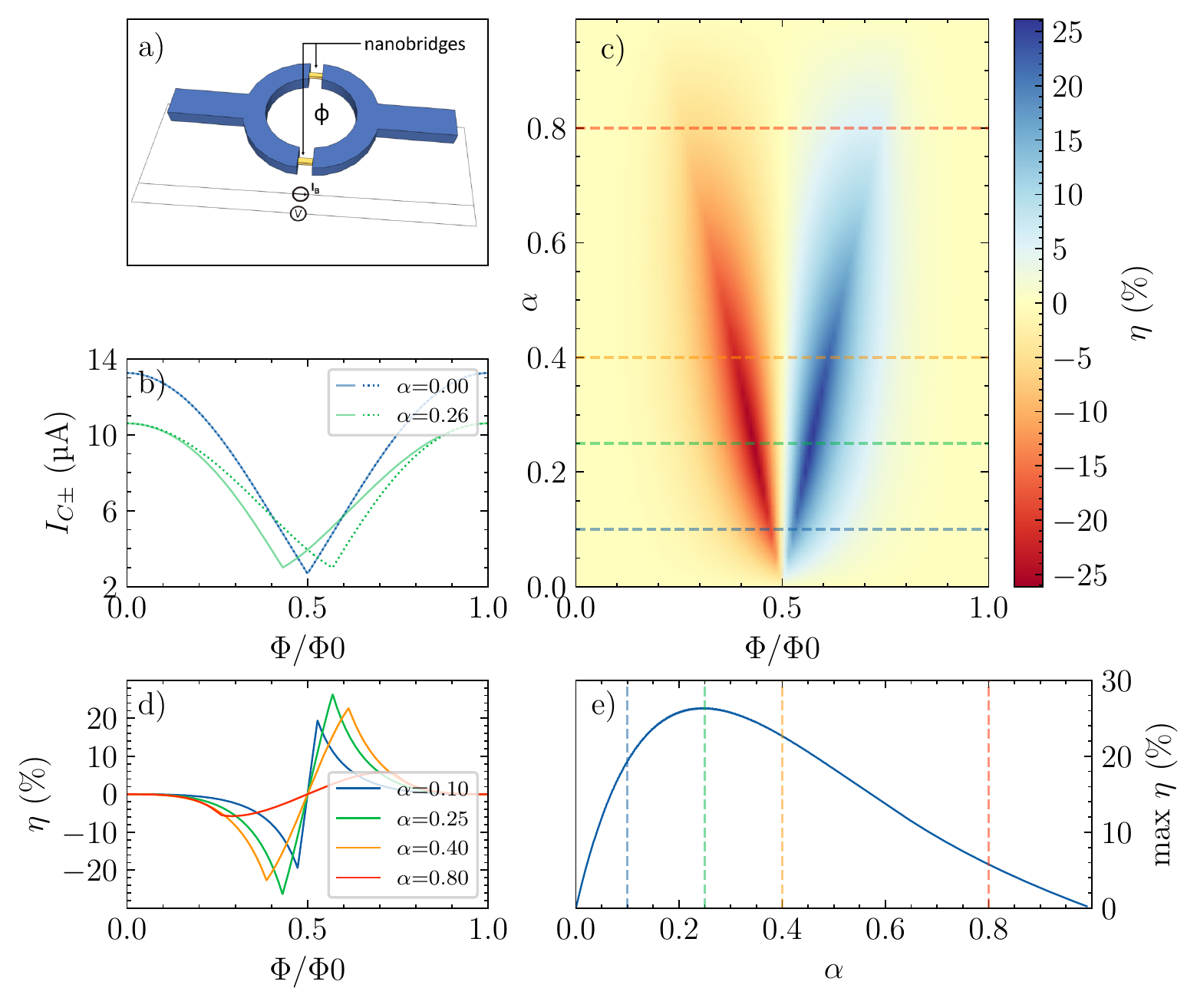}
\caption{(a) 3D model of our device. The dc-SQUID is composed of a thick Al superconducting loop (blue) embedding two short and thin Al superconducting nanowires (yellow) of length $l$. It is also shown the schematic electronic diagram used for the 4-wire measurement. $I_B$ denotes the biasing current, $V$ the voltage drop occurring across the interferometer, and $\Phi$ the external magnetic flux.
(b) Plot of $I_+$ and $I_-$ (respectively dashed and solid lines) as a function of the magnetic flux piercing the loop. The curves are calculated for two different values of the asymmetry parameter, i.e., $\alpha=0$ (blue) and $\alpha=0.25$ (green). We note that the minimum requirement for the interferometer to show the JDE is to have $\alpha\neq0$. (c) Contour plot of the rectification efficiency $\eta$ as a function of $\alpha$ and $\Phi$. The dashed lines indicate the cuts plotted in (d). (d) Cuts of contour plot (c) for different values of $\alpha$. It is clearly visible that the maximum rectification is not monotonic with growing $\alpha$, but it rather has a maximum for $\alpha\simeq0.25$, corresponding to a rectification of $\eta\simeq 26\%$. (e) Maximum rectification as a function of the  asymmetry coefficient $\alpha$. The vertical dashed lines indicate the maximum values extracted from (d).
All the curves in the plots are calculated using Eq. \ref{eq:KO_T} in the case of zero loop inductance ($L=0$).}
\label{fig:theoreticalrectification}
\end{figure}
Our device implements a monolithic all-metallic dc-SQUID that does not embed any tunnel junction, and that realizes the Josephson effect simply through geometrical constrictions. The JDE is obtained only thanks  to the high-harmonic content of the current phase relation (CPR) of the weak-links, and without the need of any relevant screening from the interferometer loop. Our measurements highlight non-hysteretic critical current vs magnetic flux characteristics characterized by a rectification efficiency as large as  $\sim 20\%$ that persists unaltered in all the explored temperature range, which goes from a base temperature of $30$ \si{\milli\kelvin} up to $800$ \si{\milli\kelvin}.\\

It has been shown in \cite{vijay2010,levensonfalk2016} that the CPR of \textit{short} diffusive metallic 3D Dayem bridges, where short means that the geometrical length ($l$) of the bridge is $l\lesssim 3.5\xi_0$ \cite{likharev1979}, with $\xi_0$ is the superconducting coherence length, can be well approximated by the Kulik - Omel'yanchuk (KO-1) relation \cite{kulik1975}

\begin{eqnarray}
I_j(\varphi,T)=\frac{\pi\Delta(T)}{eR}\cos(\varphi/2)\int_{\Delta(T)\cos(\varphi/2)}^{\Delta(T)} \frac{\tanh(\epsilon/k_BT)}{\sqrt{\epsilon^2-\Delta^2(T)\cos(\varphi/2)^2}} \,d\epsilon,
\label{eq:KO_T}
\end{eqnarray}

where $\Delta(T)$ is the temperature dependent superconducting energy gap, $e$ is the electron charge, $R$ is the normal-state resistance of the weak link, $\varphi$ is the phase drop across the weak link, and $k_B$ is the Boltzmann constant. 
By computing the CPR of a dc-SQUID [see Fig. \ref{fig:theoreticalrectification}(a)] where the two weak links respond to Eq. \ref{eq:KO_T} one obtains $I_{tot}(\varphi_1,\varphi_2,T)=I_{j,1}(\varphi_1,T)+I_{j,2}(\varphi_2,T)$, where $I_{j,n}$ is the CPR of the \textit{n}-th weak link, while $\varphi_n$ is the relative phase drop. 
Given the loop configuration of the device, the two phase drops are not independent of each other, but are in fact correlated by the flux quantization relation $\varphi_2-\varphi_1=\varphi_e-(2\pi/\Phi_0)LI_{cir}(\varphi_1,\varphi_2,T)$, where $\varphi_e=2\pi(\Phi_e/\Phi_0)$ is the phase drop induced by the external magnetic flux threading the loop $\Phi_e$, $L$ is the loop self-inductance, $I_{cir}(\varphi_1,\varphi_2,T)=I_{j,1}(\varphi_1,T)-I_{j,2}(\varphi_2,T)$ is the circulating current in the loop, and $\Phi_0$ is the flux quantum. The flux quantization relation is expressed in a transcendental form that can be solved numerically and allows us to find the phase drop of one of the two junctions with respect to the other (which we indicate as the free phase $\varphi_0$) and $\varphi_e$. In order to obtain the critical current of the dc-SQUID one then needs to find the maximum and minimum values of $I_{tot}(\varphi_0,\varphi_e,T)$ with respect to $\varphi_0$. The absolute maxima are called $I_{+}$ while the absolute minima $I_{-}$, and correspond to the highest superconducting current that the dc-SQUID can carry in the positive or negative direction, respectively. If $I_{+}$ and $I_{-}$ have a different absolute value we are in the presence of finite supercurrent rectification, which can be quantified by the rectification coefficient $\eta=(I_{+}-I_{-})/(I_{+}+I_{-})$. Moreover, we introduce the asymmetry coefficient $\alpha=(R_2-R_1)/(R_2+R_1)$, where $R_{1,2}$ are the normal-state resistances of the two weak links.\\

By using the above definitions  in the low temperature limit $T\ll T_c$ we show some theoretical results in Fig. \ref{fig:theoreticalrectification}, which are obtained in the limit of negligible loop inductance, i.e., $L=0$. This limit neglects the screening current of the dc-SQUID, thereby providing a simple analytical relation between the phase drops, $\varphi_2-\varphi_1=\varphi_e=2\pi(\Phi_e/\Phi_0)$. In (b) one can see $I_+$ and $I_-$ as a function of $\Phi_e=\Phi$ plotted for two different values of $\alpha$, namely $\alpha=0$, so the case of a perfectly symmetric dc-SQUID, and $\alpha=0.25$, hence the theoretical value for which the rectification is maximum. The JDE is revealed by the fact that $I_+$ and $I_-$ do not overlap. (c) shows $\eta$ as a function of the magnetic flux threading the loop and $\alpha$. Here we see that the rectification capability of this device is strongly dependent on the asymmetry between the two weak links, which determines both the height and position of the maximum value of $\eta$ of the $\eta(\Phi)$ characteristics. This fact is highlighted in (d), where we show some cuts of (c) plotting the rectification as a function of the magnetic flux for different values of $\alpha$. We see that the maximum rectification moves away from half flux quantum when $\alpha$ grows, and its absolute value has an absolute maximum for $\alpha\simeq0.25$. In (e) we plot the maximum rectification coefficient as a function of $\alpha$. For $\alpha=0$, hence in the case of a perfectly balanced interferometer, we have no rectification. By growing the asymmetry between the bridges we observe an increasing $\eta$, that saturates at a maximum theoretical value of $\eta\simeq 26$\% for $\alpha\simeq 0.25$. Further increasing $\alpha$ the rectification efficiency starts to decrease, vanishing for $\alpha>0.9$. This latter feature is the indication that for very large asymmetries the interference between the two CPRs becomes negligible because one supercurrent is just a small perturbation of the other. This simple model reveals that a dc-SQUID with negligible loop inductance embedding KO-1-like weak links admits a sizable supercurrent diode effect, and we stress that this is due to the nonlinearity of its CPR only. The latter can be an interesting aspect for technological purposes. Indeed achieving a sizable rectification without the need of a relevant loop inductance allows a significant downsizing of the whole device, making it a feasible candidate for on chip integration.\\

\begin{figure}
\includegraphics{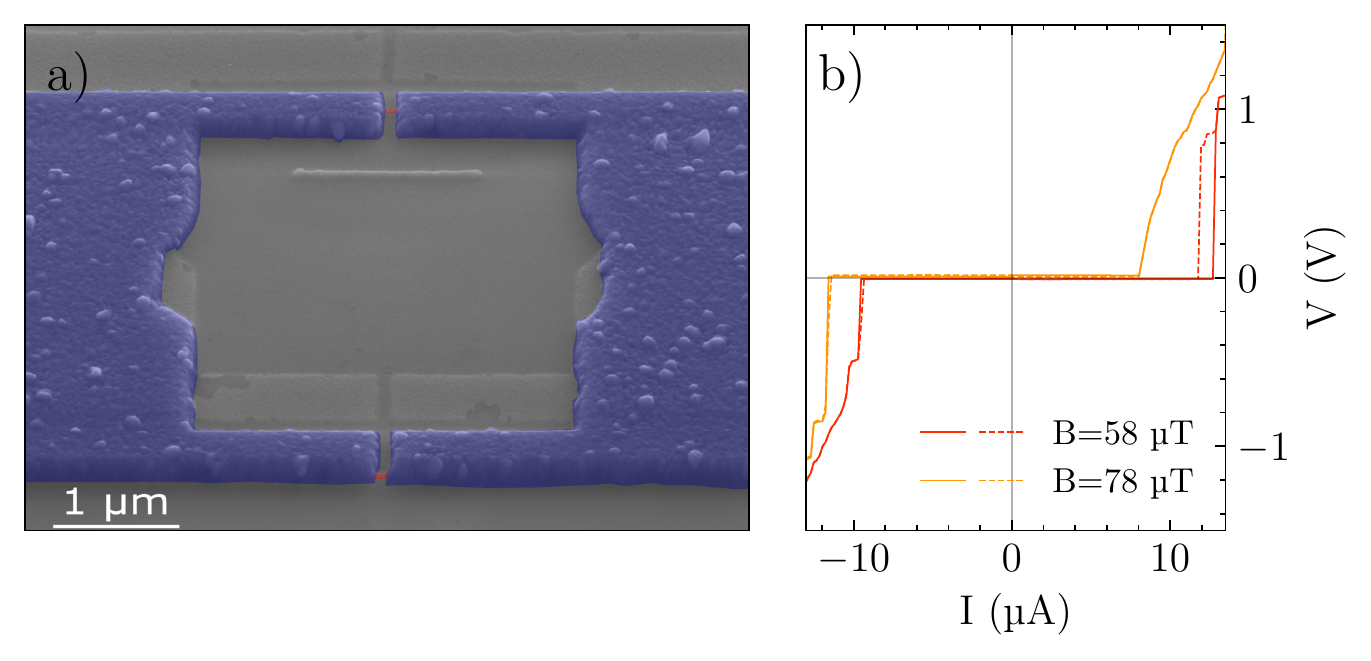}
\caption{(a) Scanning electron micrograph of the device in false colors. Aluminium nanowires (red), are $10$ \si{\nano\metre} thick and about $60$ \si{\nano\metre} long; the aluminium loop (blue) is 100 \si{\nano\metre} thick. The loop spans a surface $\approx 9$ \si{\square\micro\metre}. (b)  Current vs voltage characteristics taken at two different magnetic fields values at 30mK.\\}
\label{fig:device} 
\end{figure}

With this in hand, we show in Fig. \ref{fig:device} (a) a pseudo-color scanning electron micrograph of our superconducting interferometer based on 3D Dayem nanobridges. The dc-SQUID is composed of a $100$-\si{\nano\metre}-thick Al superconducting loop (blue) embedding two short nanowires of thickness $10$ \si{\nano\metre} (red). The loop has a square shape with an edge of $3$ \si{\micro\metre} and it is interrupted by two gaps of approximately $l\simeq 60$ \si{\nano\metre} in which the two Al nanowires of $70$ \si{\nano\metre} width are placed. The lithographic mask for the loop plus wires structure was created using a single step of electron beam lithography on a double-layer resist substrate, composed of PMMA A4 on top of MMA EL9. The Al deposition of the whole structure was realized in an UHV electron-beam evaporator through shadow-mask evaporation using two different angles. 
In the first evaporation, we deposited the thin nanowires at a tilting angle of $20$°, then rotating the sample holder to $0$° angle we evaporated the second thicker layer to realize the superconducting loop and leads. 
By performing electromagnetic simulations on the geometry of the loop we estimate $\sim 12$ \si{\pico\henry} as the value of its geometrical inductance. 
Given the thickness of the Al film we consider negligible the kinetic component of the latter. 
The dc-SQUID characterization was performed in a dilution refrigerator with a base temperature of $30$ \si{\milli\kelvin}, by collecting the $IV$ curves through a 4-wire technique. 
Figure \ref{fig:device} (b) shows two $IV$ curves measured at two different  magnetic field values, respectively $58$ \si{\micro\tesla} and $78$ \si{\micro\tesla}. It is interesting to note that, depending on the applied magnetic field, the $IV$ characteristics lose completely their hysteretic behavior, bringing the retrapping current to coincide with the switching current and lowering the characteristic voltage. From the $IV$ curves we extracted the normal-state resistance of the dc-SQUID, which corresponds to $R\simeq 26$ \si{\ohm}. We can then estimate the superconducting coherence length \cite{tinkham1975} in the bridges, $\xi_0=\sqrt{ h l/2\pi Rwte^2\Delta_0N_F}\simeq 32$ \si{\nano\metre}, with $w$ and $t$ the nanobridge width and thickness, respectively, $h$ the Plank's constant, $\Delta_0=1.764k_BT_c$ the superconducting gap at zero temperature, $N_F=2.17\cdot10^{47}$ \si{\joule^{-1}\metre^{-3}} the density of states at the Fermi level of Al, $k_B$ the Boltzmann constant, and $Tc\simeq 1.35$ \si{\kelvin} the Dayem bridge critical temperature. Since $l\lesssim 3.5\xi_0$, this value sets the frame of the short/intermediate length junction limit, thereby in the validity regime of Eq. \ref{eq:KO_T}.\\

\begin{figure*}
\includegraphics{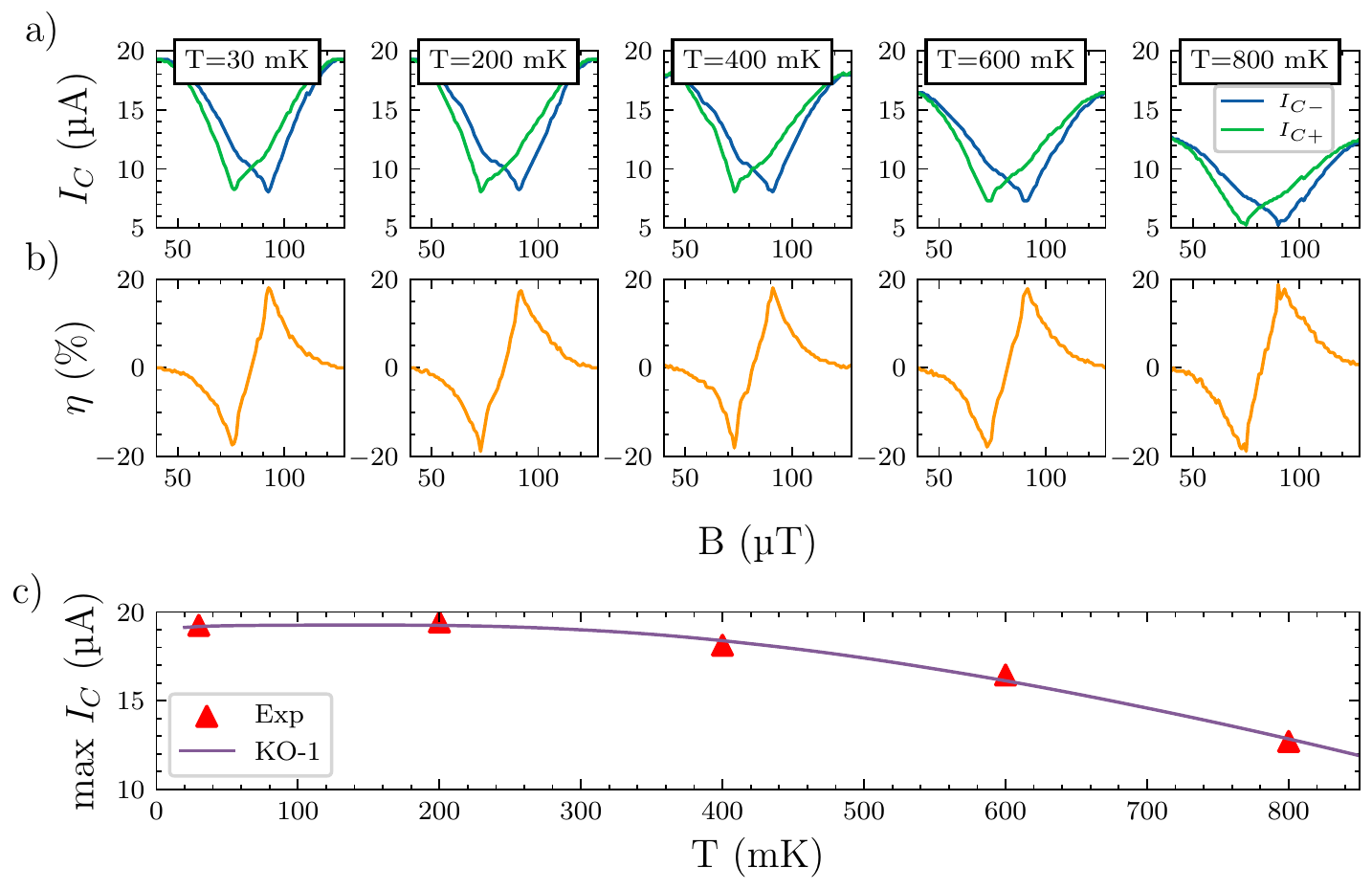}
\caption{\label{fig:temperature_characterization} (a) $I_+$ and $I_-$ as a function of magnetic field for different bath 
temperatures. (b) Rectification efficiency $\eta$ as a function of magnetic field for the same  temperatures as in panel (a). (c) Maximum value of the  critical current as a function of temperature along with the theoretical fit using Eq. \ref{eq:KO_T}.}
\end{figure*}

Figure \ref{fig:temperature_characterization} (a) displays $I_{+}$ and $I_{-}$ as a function of the applied magnetic field measured for different temperatures of the bath. We immediately note a substantial rectification effect, that persists in all the temperature range, even if the absolute value of $I_{(+,-)}$ decreases, as expected, with temperature. This fact is confirmed by Fig. \ref{fig:temperature_characterization} (b), where it is presented the rectification $\eta$ as a function of the applied magnetic field. 
These curves are extracted from the values of $I_{(+,-)}$ respectively shown in the plots above, so in the temperature range that goes from $30$ \si{\milli\kelvin} to $800$ \si{\milli\kelvin}. 
We emphasize that the behavior of $\eta$ is only marginally affected by the temperature
in all the explored range, so up to $0.59T_c$, indeed the rectification efficiency profile remains almost unaltered. This is a remarkable fact, and despite $I_{(+,-)}$ are strongly affected by temperature their relative ratio remains practically constant. This feature can be exploited for technological environments where a resilience against temperature variations is required.\\
Figure \ref{fig:temperature_characterization} (c) shows how the maximum measured critical current $I_c$ changes as a function of bath temperature, and is compared with the theoretical behavior obtained using Eq. \ref{eq:KO_T} assuming as fitting parameters $T_c=1.35$K, $R=38$ \si{\ohm}, and $\alpha=0.17$. The temperature dependence well follows the KO-1 relation, thereby confirming the validity of the model also in a wider temperature range.\\

\begin{figure}
\includegraphics{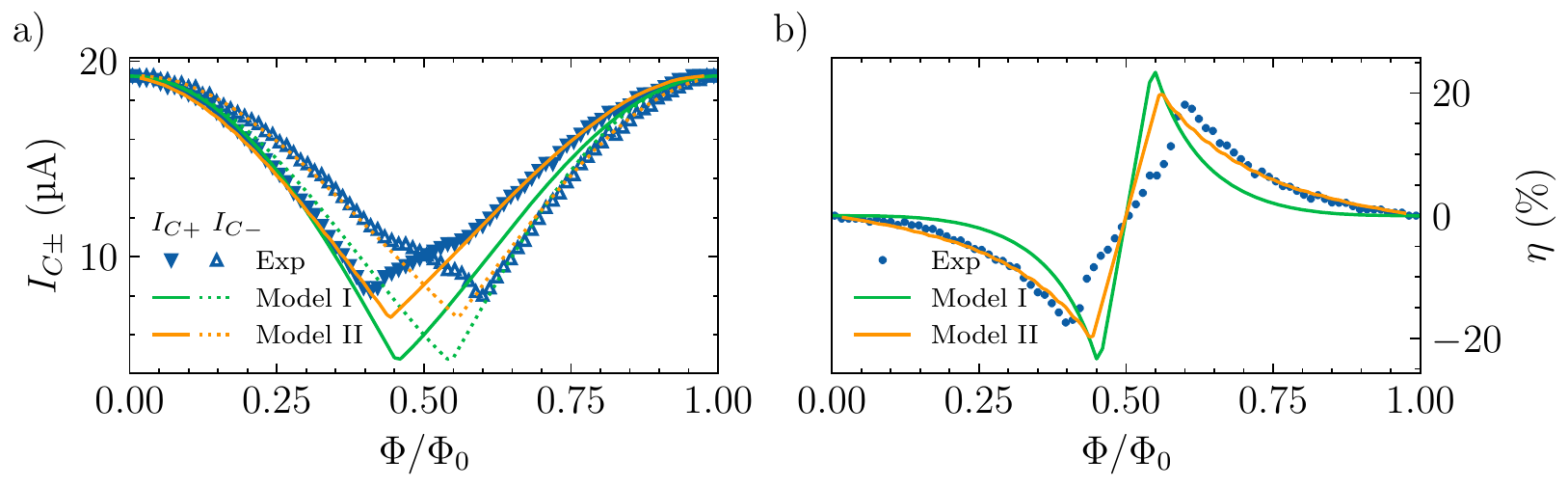}
\caption{\label{fig:rectification} (a) $I_+$ and $I_-$ as a function of magnetic flux $\Phi$ threading the loop. Triangles indicate the experimental data while the green and orange curves show the theoretical behavior for Model I and Model II, respectively. The solid curve indicates $I_+$ while the dotted curve $I_-$. 
(b) Rectification efficiency as a function of the magnetic flux, calculated using the models and data showed in (a). Dots indicate the experimental data, whereas the green and orange curves represent the theoretical rectification efficiency calculated with Model I and II, respectively.}
\end{figure}

Figure \ref{fig:rectification} (a) displays the fit of $I_{+}$ and $I_{-}$ at $30$ \si{\milli\kelvin} used to calculate the rectification at $30$\si{\milli\kelvin} [same fitting parameters of Fig. \ref{fig:temperature_characterization} (c))]. The experimental data are compared with the theoretical model in the cases of zero (Model I) and finite (Model II, $L=1 2$\si{\pico\henry}) inductance of the loop. When computing Model II we estimate the screening parameter of our dc-SQUID to be $\beta=(2I_{c}L)/\Phi_0\simeq 0.17$. The theoretical curves of Model II are in good agreement with the experimental data except in the magnetic flux range between $0.33\Phi_0$ and $0.67\Phi_0$, where we note a deeper theoretical $I_c$ suppression with respect to what we measure. A similar kind of behavior has been observed in Ref. \cite{vijay2010} while studying analogous Al devices. The authors of Ref. \cite{fominov2022} theoretically investigate the rectification capabilities of asymmetric dc-SQUIDs by computing a "minimal model" to practically achieve the JDE. From their analysis it follows that a very similar kink in the $I_c(B)$ curve occurs by introducing a different harmonic spectral content between the CPRs of the two weak links. In particular, they predict a pronounced effect by adding a second harmonic to one of the two (sinusoidal) CPRs. From this we can deduce that the theoretical curves do not follow precisely the experimental data in this specific magnetic flux region possibly because the two weak links in our device may not present exactly the same CPR, thereby including a different harmonic content. This last fact should not surprise, since the CPR of metallic weak links is known to be strongly dependent on their geometrical properties, meaning that any asymmetry between the two weak links involves a difference in their CPR as well. This issue is not taken into account in our theoretical discussion, and its inclusion goes beyond the scope of the present work. Figure \ref{fig:rectification} (b) shows  the measured rectification efficiency $\eta$ as a function of the magnetic field at $30$ \si{\milli\kelvin}. Close to zero and one flux quantum the experimental data are fitted better by Model II. This occurs because adding the effect of the screening current yields a non-negligible contribution to the rectification efficiency also for very low circulating currents, thus making non-zero the first derivative of $\eta(\Phi)$ when $\Phi$ is close to an integer flux quantum. This evidence reveals that the supercurrent diode effect can be enhanced by the finite inductance of the loop.\\
 
In summary, we showed the fabrication and characterisation of a supercurrent diode exploiting a monolithic Al dc-SQUID embedding 3D short Dayem nanobridges. The high harmonic content of the CPR typical of this interferometer allows to obtain remarkable rectification efficiencies as large as $20\%$ in a wide range of temperatures regardless of the geometrical properties of the superconducting loop. We wish to stress this last point, since it removes the miniaturization constraint due to the physical dimensions of the loop, thus allowing to push even further the minimum dimensions of this Josephson diode. In this device the JDE is obtained without the use of unconventional superconductors  or two-dimensional materials, but rather with common Al thin films, which makes this technology easy to reproduce and scalable. Moreover, higher critical temperature superconductors such as vanadium (V) \cite{spathis2011hybrid,ronzani2013micro,ronzani2014balanced} or lead (Pb) \cite{paajaste2015pb} could be easily exploited as well in order to allow the diode operation at higher bath temperatures. The general ease of fabrication makes this Josephson diode suitable for the integration with Al and Nb based superconducting electronics.\\

\begin{acknowledgement}

We acknowledge the EU’s Horizon 2020 Research and Innovation Framework Programme under Grant No. 964398 (SUPERGATE), No. 101057977 (SPECTRUM), and the PNRR MUR project PE0000023-NQSTI for partial financial support.

\end{acknowledgement}


\bibliography{achemso-demo}

\providecommand{\noopsort}[1]{}\providecommand{\singleletter}[1]{#1}%
\providecommand{\latin}[1]{#1}
\makeatletter
\providecommand{\doi}
  {\begingroup\let\do\@makeother\dospecials
  \catcode`\{=1 \catcode`\}=2 \doi@aux}
\providecommand{\doi@aux}[1]{\endgroup\texttt{#1}}
\makeatother
\providecommand*\mcitethebibliography{\thebibliography}
\csname @ifundefined\endcsname{endmcitethebibliography}
  {\let\endmcitethebibliography\endthebibliography}{}
\begin{mcitethebibliography}{30}
\providecommand*\natexlab[1]{#1}
\providecommand*\mciteSetBstSublistMode[1]{}
\providecommand*\mciteSetBstMaxWidthForm[2]{}
\providecommand*\mciteBstWouldAddEndPuncttrue
  {\def\EndOfBibitem{\unskip.}}
\providecommand*\mciteBstWouldAddEndPunctfalse
  {\let\EndOfBibitem\relax}
\providecommand*\mciteSetBstMidEndSepPunct[3]{}
\providecommand*\mciteSetBstSublistLabelBeginEnd[3]{}
\providecommand*\EndOfBibitem{}
\mciteSetBstSublistMode{f}
\mciteSetBstMaxWidthForm{subitem}{(\alph{mcitesubitemcount})}
\mciteSetBstSublistLabelBeginEnd
  {\mcitemaxwidthsubitemform\space}
  {\relax}
  {\relax}

\bibitem[Wakatsuni \latin{et~al.}(2017)Wakatsuni, Hoshino, Itahashi, Ideue,
  Ezawa, and Nagaosa]{akatsuni2017}
Wakatsuni,~R.; Hoshino,~S.; Itahashi,~Y.~M.; Ideue,~T.; Ezawa,~Y.,~M.~andIwasa;
  Nagaosa,~N. Nonreciprocal charge transport in noncentrosymmetric
  superconductors. \emph{Science Advances} \textbf{2017}, \emph{3}\relax
\mciteBstWouldAddEndPuncttrue
\mciteSetBstMidEndSepPunct{\mcitedefaultmidpunct}
{\mcitedefaultendpunct}{\mcitedefaultseppunct}\relax
\EndOfBibitem
\bibitem[Wakatsuki \latin{et~al.}(2017)Wakatsuki, Saito, Hoshino, Itahashi,
  Ideue, Ezawa, Iwasa, and Nagaosa]{wakatsuki2017}
Wakatsuki,~R.; Saito,~Y.; Hoshino,~S.; Itahashi,~Y.~M.; Ideue,~T.; Ezawa,~M.;
  Iwasa,~Y.; Nagaosa,~N. Nonreciprocal charge transport in noncentrosymmetric
  superconductors. \emph{Science Advances} \textbf{2017}, \emph{3},
  e1602390\relax
\mciteBstWouldAddEndPuncttrue
\mciteSetBstMidEndSepPunct{\mcitedefaultmidpunct}
{\mcitedefaultendpunct}{\mcitedefaultseppunct}\relax
\EndOfBibitem
\bibitem[Jiang and Hu(2022)Jiang, and Hu]{jiang2022}
Jiang,~K.; Hu,~J. Superconducting diode effects. \emph{Nature Physics}
  \textbf{2022}, \emph{18}, 1145–1146\relax
\mciteBstWouldAddEndPuncttrue
\mciteSetBstMidEndSepPunct{\mcitedefaultmidpunct}
{\mcitedefaultendpunct}{\mcitedefaultseppunct}\relax
\EndOfBibitem
\bibitem[Bauriedl \latin{et~al.}(2022)Bauriedl, Bäuml, Fuchs, Baumgartner,
  Paulik, Bauer, Lin, Lupton, Taniguchi, Watanabe, and Paradiso]{bauriedl2022}
Bauriedl,~L.; Bäuml,~C.; Fuchs,~L.; Baumgartner,~C.; Paulik,~N.; Bauer,~J.~M.;
  Lin,~K.-Q.; Lupton,~J.~M.; Taniguchi,~T.; Watanabe,~C.,~K.~Strunk;
  Paradiso,~N. Supercurrent diode effect and magnetochiral anisotropy in
  few-layer NbSe2. \emph{Nature Communications} \textbf{2022}, \emph{13},
  4266\relax
\mciteBstWouldAddEndPuncttrue
\mciteSetBstMidEndSepPunct{\mcitedefaultmidpunct}
{\mcitedefaultendpunct}{\mcitedefaultseppunct}\relax
\EndOfBibitem
\bibitem[Ando \latin{et~al.}(2020)Ando, Miyasaka, Li, Ishizuka, Arakawa,
  Shiota, Moriyama, Yanase, and Yono]{ando2020}
Ando,~F.; Miyasaka,~Y.; Li,~T.; Ishizuka,~J.; Arakawa,~T.; Shiota,~Y.;
  Moriyama,~T.; Yanase,~Y.; Yono,~T. Observation of superconducting diode
  effect. \emph{Nature} \textbf{2020}, \emph{584}, 373--376\relax
\mciteBstWouldAddEndPuncttrue
\mciteSetBstMidEndSepPunct{\mcitedefaultmidpunct}
{\mcitedefaultendpunct}{\mcitedefaultseppunct}\relax
\EndOfBibitem
\bibitem[Gupta \latin{et~al.}(2023)Gupta, Graziano, Pendharkar, Dong, Dempsey,
  Palmstrøm, and Pribiag]{gupta2023}
Gupta,~M.; Graziano,~G.~V.; Pendharkar,~M.; Dong,~J.~T.; Dempsey,~C.~P.;
  Palmstrøm,~C.; Pribiag,~V.~S. Gate-tunable superconducting diode effect in a
  three-terminal Josephson device. \emph{Nature Communications} \textbf{2023},
  \emph{14}, 3078\relax
\mciteBstWouldAddEndPuncttrue
\mciteSetBstMidEndSepPunct{\mcitedefaultmidpunct}
{\mcitedefaultendpunct}{\mcitedefaultseppunct}\relax
\EndOfBibitem
\bibitem[Wu \latin{et~al.}(2022)Wu, Wang, Xu, Sivakumar, Pasco, Filippozzi,
  Parkin, Zeng, McQueen, and Ali]{wu2022}
Wu,~H.; Wang,~Y.; Xu,~Y.; Sivakumar,~P.~K.; Pasco,~C.; Filippozzi,~U.;
  Parkin,~S. S.~P.; Zeng,~Y.-J.; McQueen,~T.; Ali,~M.~N. The field-free
  Josephson diode in a van der Waals heterostructure. \emph{Nature}
  \textbf{2022}, \emph{604}, 653–656\relax
\mciteBstWouldAddEndPuncttrue
\mciteSetBstMidEndSepPunct{\mcitedefaultmidpunct}
{\mcitedefaultendpunct}{\mcitedefaultseppunct}\relax
\EndOfBibitem
\bibitem[Turini \latin{et~al.}(2022)Turini, Salimian, Carrega, Iorio,
  Strambini, Giazotto, Zannier, Sorba, and Heun]{turini2022josephson}
Turini,~B.; Salimian,~S.; Carrega,~M.; Iorio,~A.; Strambini,~E.; Giazotto,~F.;
  Zannier,~V.; Sorba,~L.; Heun,~S. Josephson Diode Effect in High-Mobility InSb
  Nanoflags. \emph{Nano Letters} \textbf{2022}, \emph{22}, 8502--8508\relax
\mciteBstWouldAddEndPuncttrue
\mciteSetBstMidEndSepPunct{\mcitedefaultmidpunct}
{\mcitedefaultendpunct}{\mcitedefaultseppunct}\relax
\EndOfBibitem
\bibitem[Aladyshkin \latin{et~al.}(2009)Aladyshkin, Fritzsche, and
  Moshchalkov]{aladyshkin2009}
Aladyshkin,~A.~Y.; Fritzsche,~J.; Moshchalkov,~V.~V. Planar
  superconductor/ferromagnet hybrids: Anisotropy of resistivity induced by
  magnetic templates. \emph{Appl. Phys. Lett.} \textbf{2009}, \emph{94}\relax
\mciteBstWouldAddEndPuncttrue
\mciteSetBstMidEndSepPunct{\mcitedefaultmidpunct}
{\mcitedefaultendpunct}{\mcitedefaultseppunct}\relax
\EndOfBibitem
\bibitem[Lyu \latin{et~al.}(2021)Lyu, Jiang, Wang, Xiao, Dong, Chen,
  Milošević, Wang, Divan, Pearson, Wu, Peters, and Kwok]{lyu2021}
Lyu,~Y.-Y.; Jiang,~J.; Wang,~Y.-L.; Xiao,~Z.-L.; Dong,~S.; Chen,~Q.-H.;
  Milošević,~M.~V.; Wang,~H.; Divan,~R.; Pearson,~J.~E.; Wu,~P.;
  Peters,~F.~M.; Kwok,~W.-K. {Superconducting diode effect via conformal-mapped
  nanoholes}. \emph{Nature Communications} \textbf{2021}, \emph{12}\relax
\mciteBstWouldAddEndPuncttrue
\mciteSetBstMidEndSepPunct{\mcitedefaultmidpunct}
{\mcitedefaultendpunct}{\mcitedefaultseppunct}\relax
\EndOfBibitem
\bibitem[Satchell \latin{et~al.}(2023)Satchell, Shepley, Rosamond, and
  Burnell]{satchell2023}
Satchell,~N.; Shepley,~P.~M.; Rosamond,~M.~C.; Burnell,~G. {Supercurrent diode
  effect in thin film Nb tracks}. \emph{Journal of Applied Physics}
  \textbf{2023}, \emph{133}\relax
\mciteBstWouldAddEndPuncttrue
\mciteSetBstMidEndSepPunct{\mcitedefaultmidpunct}
{\mcitedefaultendpunct}{\mcitedefaultseppunct}\relax
\EndOfBibitem
\bibitem[Zhang \latin{et~al.}(2022)Zhang, Gu, Li, Hu, and Jiang]{zhang2022}
Zhang,~Y.; Gu,~Y.; Li,~P.; Hu,~J.; Jiang,~K. General Theory of Josephson
  Diodes. \emph{Phys. Rev. X} \textbf{2022}, \emph{12}, 041013\relax
\mciteBstWouldAddEndPuncttrue
\mciteSetBstMidEndSepPunct{\mcitedefaultmidpunct}
{\mcitedefaultendpunct}{\mcitedefaultseppunct}\relax
\EndOfBibitem
\bibitem[Baumgartner \latin{et~al.}(2022)Baumgartner, Fuchs, Costa, Reinhardt,
  Gronin, Gardner, Lindemann, Manfra, Faria~Junior, Kochan, Fabian, Paradiso,
  and Strunk]{baumgartner2022}
Baumgartner,~C.; Fuchs,~L.; Costa,~A.; Reinhardt,~S.; Gronin,~S.;
  Gardner,~G.~C.; Lindemann,~T.; Manfra,~M.~J.; Faria~Junior,~P.~E.;
  Kochan,~D.; Fabian,~J.; Paradiso,~N.; Strunk,~C. Supercurrent rectification
  and magnetochiral effects in symmetric Josephson junctions. \emph{Nature
  Nanotechnology} \textbf{2022}, \emph{17}, 628--633\relax
\mciteBstWouldAddEndPuncttrue
\mciteSetBstMidEndSepPunct{\mcitedefaultmidpunct}
{\mcitedefaultendpunct}{\mcitedefaultseppunct}\relax
\EndOfBibitem
\bibitem[Fulton \latin{et~al.}(1972)Fulton, Dunkleberger, and
  Dynes]{fulton1972}
Fulton,~T.~A.; Dunkleberger,~L.~N.; Dynes,~R.~C. Quantum Interference
  Properties of Double Josephson Junctions. \emph{Phys. Rev. B} \textbf{1972},
  \emph{6}, 855--875\relax
\mciteBstWouldAddEndPuncttrue
\mciteSetBstMidEndSepPunct{\mcitedefaultmidpunct}
{\mcitedefaultendpunct}{\mcitedefaultseppunct}\relax
\EndOfBibitem
\bibitem[Souto \latin{et~al.}(2022)Souto, Leijnse, and Schrade]{souto2022}
Souto,~R.~S.; Leijnse,~M.; Schrade,~C. Josephson Diode Effect in Supercurrent
  Interferometers. \emph{Phys. Rev. Lett.} \textbf{2022}, \emph{129},
  267702\relax
\mciteBstWouldAddEndPuncttrue
\mciteSetBstMidEndSepPunct{\mcitedefaultmidpunct}
{\mcitedefaultendpunct}{\mcitedefaultseppunct}\relax
\EndOfBibitem
\bibitem[Paolucci \latin{et~al.}(2023)Paolucci, De~Simoni, and
  Giazotto]{paolucci2023}
Paolucci,~F.; De~Simoni,~G.; Giazotto,~F. A gate- and flux-controlled
  supercurrent diode effect. \emph{Appl. Phys. Lett.} \textbf{2023},
  \emph{122}\relax
\mciteBstWouldAddEndPuncttrue
\mciteSetBstMidEndSepPunct{\mcitedefaultmidpunct}
{\mcitedefaultendpunct}{\mcitedefaultseppunct}\relax
\EndOfBibitem
\bibitem[Levenson-Falk \latin{et~al.}(2011)Levenson-Falk, Vijay, and
  Siddiqi]{levensonfalk2011}
Levenson-Falk,~E.~M.; Vijay,~R.; Siddiqi,~I. Nonlinear microwave response of
  aluminum weak-link Josephson oscillators. \emph{Appl. Phys. Lett.}
  \textbf{2011}, \emph{98}\relax
\mciteBstWouldAddEndPuncttrue
\mciteSetBstMidEndSepPunct{\mcitedefaultmidpunct}
{\mcitedefaultendpunct}{\mcitedefaultseppunct}\relax
\EndOfBibitem
\bibitem[Bouchiat \latin{et~al.}(2001)Bouchiat, Faucher, Thirion, Wernsdorfer,
  Fournier, and Pannetier]{bouchiat2001}
Bouchiat,~V.; Faucher,~M.; Thirion,~C.; Wernsdorfer,~W.; Fournier,~T.;
  Pannetier,~B. Josephson junctions and superconducting quantum interference
  devices made by local oxidation of niobium ultrathin films. \emph{Appl. Phys.
  Lett.} \textbf{2001}, \emph{79}\relax
\mciteBstWouldAddEndPuncttrue
\mciteSetBstMidEndSepPunct{\mcitedefaultmidpunct}
{\mcitedefaultendpunct}{\mcitedefaultseppunct}\relax
\EndOfBibitem
\bibitem[Margineda \latin{et~al.}(2023)Margineda, Crippa, Strambini, Fukaya,
  Mercaldo, and Giazotto]{margineda2023sign}
Margineda,~D.; Crippa,~A.; Strambini,~E.; Fukaya,~Y.; Mercaldo,~M.~T.;
  Giazotto,~F. Sign reversal diode effect in superconducting Dayem nanobridges.
  \emph{arXiv preprint arXiv:2306.00193} \textbf{2023}, \relax
\mciteBstWouldAddEndPunctfalse
\mciteSetBstMidEndSepPunct{\mcitedefaultmidpunct}
{}{\mcitedefaultseppunct}\relax
\EndOfBibitem
\bibitem[Vijay \latin{et~al.}(2010)Vijay, Levenson-Falk, Slichter, and
  Siddiqi]{vijay2010}
Vijay,~R.; Levenson-Falk,~E.~M.; Slichter,~D.~H.; Siddiqi,~I. {Approaching
  ideal weak link behavior with three dimensional aluminum nanobridges}.
  \emph{Applied Physics Letters} \textbf{2010}, \emph{96}, 223112\relax
\mciteBstWouldAddEndPuncttrue
\mciteSetBstMidEndSepPunct{\mcitedefaultmidpunct}
{\mcitedefaultendpunct}{\mcitedefaultseppunct}\relax
\EndOfBibitem
\bibitem[Levenson-Falk \latin{et~al.}(2016)Levenson-Falk, Antler, and
  Siddiqi]{levensonfalk2016}
Levenson-Falk,~E.~M.; Antler,~N.; Siddiqi,~I. Dispersive nanoSQUID
  magnetometry. \emph{Superconductor Science and Technology} \textbf{2016},
  \emph{29}, 113003\relax
\mciteBstWouldAddEndPuncttrue
\mciteSetBstMidEndSepPunct{\mcitedefaultmidpunct}
{\mcitedefaultendpunct}{\mcitedefaultseppunct}\relax
\EndOfBibitem
\bibitem[Likharev(1979)]{likharev1979}
Likharev,~K.~K. Superconducting weak links. \emph{Rev. Mod. Phys.}
  \textbf{1979}, \emph{51}, 101--159\relax
\mciteBstWouldAddEndPuncttrue
\mciteSetBstMidEndSepPunct{\mcitedefaultmidpunct}
{\mcitedefaultendpunct}{\mcitedefaultseppunct}\relax
\EndOfBibitem
\bibitem[Kulik and Omel'yanchuk(1975)Kulik, and Omel'yanchuk]{kulik1975}
Kulik,~I.~O.; Omel'yanchuk,~A.~N. Contribution to the microscopic theory of the
  Josephson effect in superconducting bridges. \emph{JETP Lett. (USSR) (Engl.
  Transl.), v. 21, no. 4, pp. 96-97} \textbf{1975}, \relax
\mciteBstWouldAddEndPunctfalse
\mciteSetBstMidEndSepPunct{\mcitedefaultmidpunct}
{}{\mcitedefaultseppunct}\relax
\EndOfBibitem
\bibitem[Tinkham(1975)]{tinkham1975}
Tinkham,~M. \emph{Introduction to superconductivity}; McGraw-Hill, 1975\relax
\mciteBstWouldAddEndPuncttrue
\mciteSetBstMidEndSepPunct{\mcitedefaultmidpunct}
{\mcitedefaultendpunct}{\mcitedefaultseppunct}\relax
\EndOfBibitem
\bibitem[Fominov and Mikhailov(2022)Fominov, and Mikhailov]{fominov2022}
Fominov,~Y.~V.; Mikhailov,~D.~S. Asymmetric higher-harmonic SQUID as a
  Josephson diode. \emph{Phys. Rev. B} \textbf{2022}, \emph{106}, 134514\relax
\mciteBstWouldAddEndPuncttrue
\mciteSetBstMidEndSepPunct{\mcitedefaultmidpunct}
{\mcitedefaultendpunct}{\mcitedefaultseppunct}\relax
\EndOfBibitem
\bibitem[Spathis \latin{et~al.}(2011)Spathis, Biswas, Roddaro, Sorba, Giazotto,
  and Beltram]{spathis2011hybrid}
Spathis,~P.; Biswas,~S.; Roddaro,~S.; Sorba,~L.; Giazotto,~F.; Beltram,~F.
  Hybrid InAs nanowire--vanadium proximity SQUID. \emph{Nanotechnology}
  \textbf{2011}, \emph{22}, 105201\relax
\mciteBstWouldAddEndPuncttrue
\mciteSetBstMidEndSepPunct{\mcitedefaultmidpunct}
{\mcitedefaultendpunct}{\mcitedefaultseppunct}\relax
\EndOfBibitem
\bibitem[Ronzani \latin{et~al.}(2013)Ronzani, Baillergeau, Altimiras, and
  Giazotto]{ronzani2013micro}
Ronzani,~A.; Baillergeau,~M.; Altimiras,~C.; Giazotto,~F. Micro-superconducting
  quantum interference devices based on V/Cu/V Josephson nanojunctions.
  \emph{Applied Physics Letters} \textbf{2013}, \emph{103}, 052603\relax
\mciteBstWouldAddEndPuncttrue
\mciteSetBstMidEndSepPunct{\mcitedefaultmidpunct}
{\mcitedefaultendpunct}{\mcitedefaultseppunct}\relax
\EndOfBibitem
\bibitem[Ronzani \latin{et~al.}(2014)Ronzani, Altimiras, and
  Giazotto]{ronzani2014balanced}
Ronzani,~A.; Altimiras,~C.; Giazotto,~F. Balanced double-loop mesoscopic
  interferometer based on Josephson proximity nanojunctions. \emph{Applied
  Physics Letters} \textbf{2014}, \emph{104}, 032601\relax
\mciteBstWouldAddEndPuncttrue
\mciteSetBstMidEndSepPunct{\mcitedefaultmidpunct}
{\mcitedefaultendpunct}{\mcitedefaultseppunct}\relax
\EndOfBibitem
\bibitem[Paajaste \latin{et~al.}(2015)Paajaste, Amado, Roddaro, Bergeret,
  Ercolani, Sorba, and Giazotto]{paajaste2015pb}
Paajaste,~J.; Amado,~M.; Roddaro,~S.; Bergeret,~F.; Ercolani,~D.; Sorba,~L.;
  Giazotto,~F. Pb/InAs nanowire Josephson junction with high critical current
  and magnetic flux focusing. \emph{Nano letters} \textbf{2015}, \emph{15},
  1803--1808\relax
\mciteBstWouldAddEndPuncttrue
\mciteSetBstMidEndSepPunct{\mcitedefaultmidpunct}
{\mcitedefaultendpunct}{\mcitedefaultseppunct}\relax
\EndOfBibitem
\end{mcitethebibliography}

\end{document}